# Soliton Shedding from Airy Pulses in a Highly Dispersive and Nonlinear Medium

DEEPENDRA SINGH GAUR, ANKIT PUROHIT, AND AKHILESH KUMAR MISHRA[*]

*Department of Physics; Indian Institute of Technology Roorkee, Roorkee-247667, Uttarakhand, India*
*\*akhilesh.mishra@ph.iitr.ac.in*

**Abstract:** We present a numerical investigation of the propagation dynamics of a truncated Airy pulse in a highly dispersive and nonlinear medium by employing the split-step Fourier transform method and look, in particular, into the effects of fourth order dispersion (FOD) and cubic-quintic-septic nonlinearity on pulse evolution. Presence of FOD cancels the Airy pulse's self-acceleration along with eclipsing the oscillatory tail during propagation in the linear regime. Further, we observe soliton shedding at low input pulse power in the presence of cubic and quintic nonlinearity and negative FOD. The emergent soliton exhibits temporal shift and the direction and the extent of the shift depend upon the strengths of cubic and quintic nonlinearities. In the presence of anomalous GVD with negative FOD, soliton shedding is observed at relatively high input pulse power. The strengths of GVD and nonlinearity play a vital role in the temporal shifting of emergent soliton. Furthermore, we have explored the effects of septic nonlinearity on soliton shedding in different scenarios of nonlinearity and dispersion.

**1. Introduction**

Airy function, as a solution to potential free Schrödinger equation, had been investigated first by M.V. Berry and Balazs in 1979 in the context of quantum mechanics. The theoretical study demonstrated its remarkable features, such as self-acceleration, non-spreading, and self-healing [1]. Since the probability density of this Airy wave-packet remains constant, the wave packet was inherently nondiffracting in nature. This also indicates the infinite power this wave-packet encompasses, therefore its synthesis was difficult for practical applications. But the wave-packet has attracted a great deal attention from the beginning due to its above mentioned remarkable features. Mathematically the Airy wave-packet is represented as [1]

$$\psi(x,0) = Ai(\alpha x/\hbar^{\frac{2}{3}}) \qquad (1)$$

where Ai represents the Airy function and $\alpha$ is an arbitrary constant.

In the framework of optics, by exploiting analogy between the optical paraxial wave equation and the potential-free Schrödinger equation, G. A. Siviloglou and D. N. Christodoulides first theoretically predicted and later made the first experimental observation of Airy beams [2]. The experimental realization was made possible by the Gaussian beam modulation with the cubic spectral phase [3]. As first noted in [1], in one dimension (1D), this Airy packet happens to be unique, e.g., it is the only nontrivial solution (apart from a plane wave) that remains invariant with time. These beams, in contrast to the already known families of non-diffracting fields, do not result from conical superposition and, as stated before, are possible even in 1D. The finite energy airy beam (FEAB) intensity follows parabolic path even in the absence of external force [4], which attracted considerable attention in the field of optical trapping and manipulation [5,6], laser filamentation [7,8], beam focusing, and nonlinear optics [9,10].

More recently, using the analogy between "diffraction in space" and "dispersion in time", the concept of the truncated Airy pulse has been introduced. Notice that the temporal attributes are the direct translation of the spatial Airy beam features. Intensity maxima of Airy pulse follow a ballistic trajectory, which is a manifestation of acceleration in the retarded time frame [11]. Because of this, dynamic propagation of Airy pulse stimulates great research interest in linear dispersive as well as in nonlinear media.

Gaussian pulse propagating near zero-dispersion wavelength, in a fiber generates the Airy pulse under the influence of third order dispersion (TOD) [3]. Apart from this, there are several methods to produce the Airy pulse, beam shaping technique is one of them, in which a cubic spectral phase is imposed on a Gaussian pulse through the computer-generated device spatial light modulator (SLM). The output is given by the product of the exponential decay and Airy function [12].

The propagation of the Airy pulse has been extensively studied in linear as well as in the nonlinear regime, in which the combined effect of GVD along with quadratic and Kerr nonlinearity has been analyzed [13-15]. In particular, soliton shedding [16], nonlinear spectral reshaping [17], inversion, and tight focusing [18] have gained special attention. Furthermore, the pulse propagation dynamics in the context of other higher-order effects such as self-steepening, Raman effect is studied more recently [19]. The observation of spatiotemporal optical soliton in cubic quintic septic nonlinearity is also reported for gaussian pulses [20]. However, the effects of higher-order nonlinearities including quintic and septic are not studied well for Airy pulse.

Several studies suggest that effects of GVD and TOD can be minimized under certain conditions[21-23]. In such cases, propagation of high-intensity ultrashort pulse can be accurately described by considering higher-order dispersion effects like FOD. The solitary wave was found as a solution to the Nonlinear Schrodinger equation (NLSE) [24] in the presence of higher-order dispersion and nonlinear effects. Besides, more recently the pure-quartic soliton has been observed in the presence of negative FOD and positive Kerr nonlinearity [25-27]. In this paper, we examine the truncated Airy pulse propagation in a medium where FOD and cubic, quintic and septic nonlinearities play crucial roles. Apart from that, propagation was also investigated with different possible combinations of dispersion present in the medium.

## 2. Theoretical Model

In the presence of higher-order dispersion and cubic-quintic-septic nonlinearity, the propagation dynamics of an ultrashort pulse can be described by the NLSE [24,26,28].

$$i\frac{\partial A}{\partial \xi} - \frac{\beta_2}{2}\frac{\partial^2 A}{\partial T^2} - \frac{i\beta_3}{6}\frac{\partial^3 A}{\partial T^3} + \frac{\beta_4}{24}\frac{\partial^4 A}{\partial T^4} + \gamma_1(\omega_0)|A|^2 A + \gamma_2(\omega_0)|A|^4 A + \gamma_3(\omega_0)|A|^6 A = 0 \quad (2)$$

where A is the slowly varying envelope of the field and $T = t - \frac{z}{v_g}$ is the time in the pulse frame, where $t$ is real time and $v_g$ is the group velocity. $\beta_2$, $\beta_3$ and $\beta_4$ are the GVD, TOD and FOD coefficient respectively. $\gamma_1$, $\gamma_2$ and $\gamma_3$ are the nonlinear coefficient and are related to $n_2$, $n_4$ and $n_6$ with $\gamma_1 = \frac{n_2(\omega_0)\,\omega_0}{c\,A_{eff}}$, $\gamma_2 = \frac{n_4(\omega_0)\,\omega_0}{c\,A_{eff1}}$ and $\gamma_3 = \frac{n_6(\omega_0)\omega_0}{c\,A_{eff2}}$ respectively, where c is the speed of light, $\omega_0$ is the carrier frequency and $A_{eff}$ is the effective mode area of guiding medium and $A_{eff1} = \frac{3}{4}A_{eff}$, $A_{eff2} = \frac{1}{2}A_{eff}$ [28].

To examine the propagation of truncated Airy pulses, equation (2) is solved numerically using the split-step Fourier transform method (SSFT). Dimensionless form of equation (2) is used in numerical simulation. Therefore we rewrite the equation (2) in the form of dimensionless coordinates. The normalized parameters are defined as $\tau = \frac{T}{T_0}$, $\psi = \frac{A}{\sqrt{P_0}}$, $Z = \frac{\xi}{L_{D4}}$, where $T_0$ is the pulse width and $P_0$ is the peak power of the pulse. The characteristic length $L_{Dm} = \frac{T_0^m}{|\beta_m|}$ $(m = 2, 3, 4..)$ and $L_{NL1} = \frac{1}{\gamma_1 P_0}$, $L_{NL2} = \frac{1}{\gamma_2 P_0^2}$ and $L_{NL3} = \frac{1}{\gamma_3 P_0^3}$ are the $m^{th}$ order dispersion length and cubic, quintic and septic nonlinear lengths respectively.

So equation (2) takes the following form

$$\frac{\partial \psi}{\partial Z} = -i\frac{\delta_2}{2}\frac{\partial^2 \psi}{\partial \tau^2} + \frac{\delta_3}{6}\frac{\partial^3 \psi}{\partial \tau^3} + i\frac{\delta_4}{24}\frac{\partial^4 \psi}{\partial \tau^4} + iN|\psi|^2\psi + i\sigma|\psi|^4\psi + i\theta|\psi|^6\psi \quad (3)$$

where $\delta_4 = sgn(\beta_4)$, $\delta_3 = sgn(\beta_3)\frac{L_{D4}}{L_{D3}} = \frac{T_0 \beta_3}{|\beta_4|}$, $\delta_2 = sgn(\beta_2)\frac{L_{D4}}{L_{D2}} = \frac{T_0^2 \beta_2}{|\beta_4|}$, and since these parameters inversely depend on the FOD, hence it is indispensable to consider these for modeling ultra-short pulse dynamics. The parameters $N = \frac{L_{D4}}{L_{NL1}} = \frac{\gamma_1 P_0 T_0^4}{|\beta_4|}$, $\sigma = \frac{L_{D4}}{L_{NL2}} = \frac{\gamma_2 P_0^2 T_0^4}{|\beta_4|}$, and $\theta = \frac{L_{D4}}{L_{NL3}} = \frac{\gamma_3 P_0^3 T_0^4}{|\beta_4|}$ are proportional to the input pulse power.

The normalized input truncated Airy pulse profile can be written as

$$\psi(Z = 0, \tau) = k(a)\, Airy(\tau)\exp(a\tau) \quad (4)$$

where 'a' is the truncation factor ($0 < a < 1$) and $k$ is truncation factor dependent parameter, which keeps the input pulse intensity at 1 for any value of 'a'. Here, truncation factor 'a' is chosen 0.1.

### 3. Numerical Result and Discussion

#### 3.1 Effect of FOD in Linear Regime

The availability of dispersion- flattened fibers with an extremum of the GVD at the operating wavelength facilitate the propagation of optical pulses with minimized TOD induced effects [21,23]. Therefore, FOD emerges as a dominant dispersion effect and to reveal the effect of FOD on Airy pulse propagation, we selectively picked it and set all terms zero in equation (3). The corresponding numerical results are demonstrated in fig. 1(a) and 1(b), which conclude that oscillatory tail of the Airy pulse is terminated during the propagation and the asymmetric input Airy pulse reshape itself into a symmetric pulse by transferring the side lobe energy to the main lobe. Moreover, the inherent self-acceleration property of Airy pulse is also vanished. Also, the Airy pulse evolution is found to be identical irrespective of the sign of FOD term. Note that similar to GVD, FOD is an even power dispersion term in NLSE, still the self-acceleration as observed in presence of GVD, is suppressed by FOD as illustrated in fig. 1.

Additionally, we also examined Airy pulse propagation with both GVD and FOD nonzero and the corresponding numerical results are shown in fig. 2(a) and fig. 2(b). We observed that when GVD and FOD are of the same sign, pulse intensity decreases rapidly, and pulse appears to be symmetric at the output, as shown in fig. 2(b). In contrast, when FOD and GVD are of the opposite sign, pulse intensity decreases slowly and ultimately assumes a symmetric shape as demonstrated in fig. 2(a).

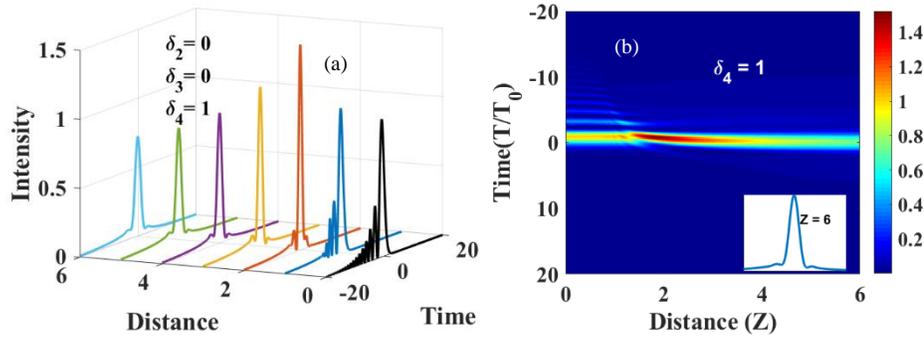

**Fig. 1.** Pulse evolution in the time domain when **(a)** only FOD is present **(b)** Contour plot of the same.

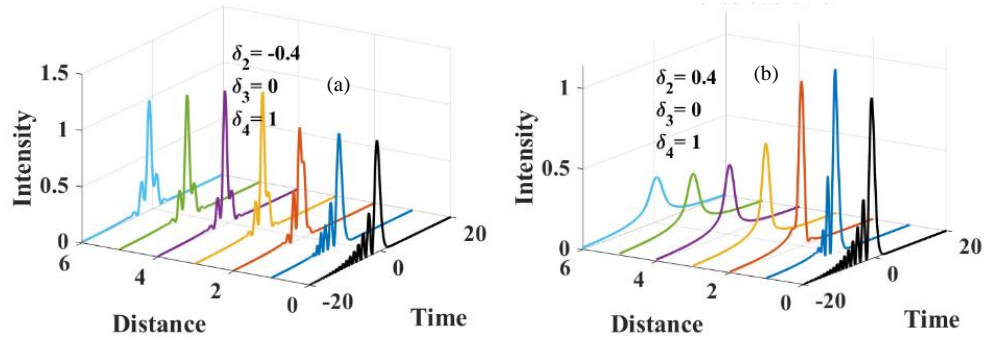

**Fig. 2.** Pulse evolution in the time domain when **(a)** GVD and FOD are of opposite in sign **(b)** GVD and FOD are of the same sign.

### 3.2 Effect of FOD and Cubic Nonlinearity

The existence of the soliton in the presence of GVD and SPM for Airy pulses has already been studied rigorously [16]. A recent observation of pure quartic soliton stimulates such investigation for Airy pulses in the presence of FOD and cubic nonlinearity [25]. Since FOD exhibits the same symmetry as GVD hence the effects of negative FOD must be identical to anomalous GVD. Therefore, in this section we study the Airy pulse propagation in presence of negative FOD and Kerr nonlinearity by varying the parameter N (and therefore the input power) defined in equation (3). The temporal evolution of the pulse for three values of N is shown in fig. 3 (a-c). Interestingly, Airy pulse intensity reaches a maximum before soliton shedding because the Airy side lobes merges to the main lobe and the peak intensity is enhanced. Propagating pulse is benefited by this additional power and therefore the soliton shedding is successfully observed even for a small value of N (= 0.2). Since, the Airy side lobe merges into the main lobe, the accelerating wavefront, resulted by the self-healing property of Airy pulse does not appear here. Hence, the background dispersive radiation is not observed [16]. For the larger value of N, intense soliton, shifted towards the leading edge, is observed as illustrated in fig. 3 (b-c). Apart from that, considerable change is noticed in the corresponding spectral evolutions, which are demonstrated in fig. 3 (d-f). In comparison to conventional solitons, these solitons are not only observed at low input pulse power but also exhibit oscillatory tails both in leading and trailing edges during the propagation, which can easily be appreciated on a log scale as depicted in fig. 4 (a).

Additionally, the intensity maximum and temporal position of corresponding emergent solitons are shown in fig. 4 (b-c). For larger N (larger input power), more intense solitons are observed.

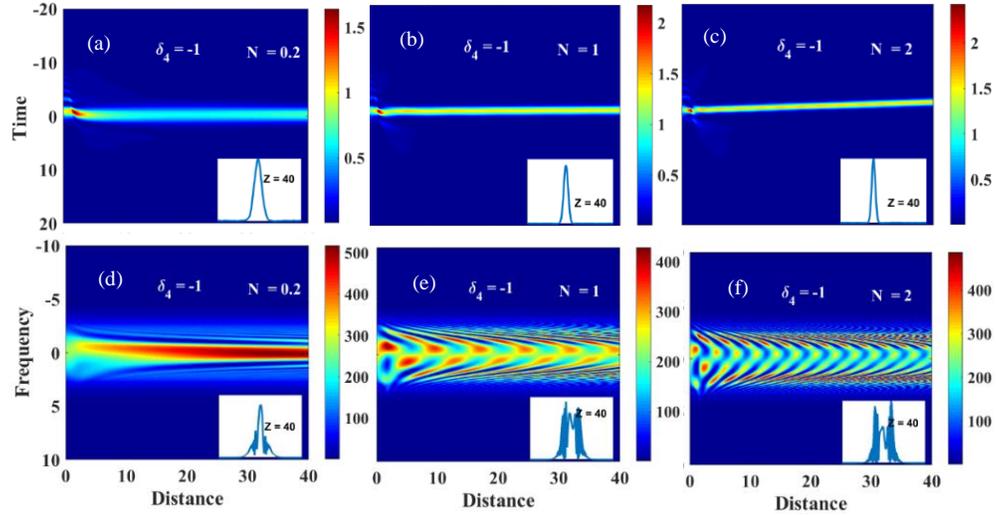

**Fig. 3.** Contour plots of temporal (upper row) and spectral (lower row) Airy pulse evolution under the influence of FOD and cubic nonlinearity. **(a), (d)** N = 0.2, **(b), (e)** N = 1, and **(c), (f)** N = 2

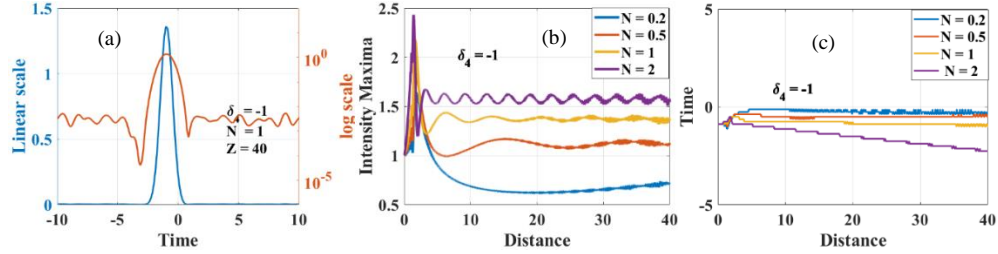

**Fig. 4. (a)** Intensity of output pulse on linear and log scales, **(b)** intensity maxima of emergent soliton with propagation distance, and **(c)** temporal position of emergent soliton with propagation distance.

Additionally, it is worth mentioning that maximum intensity of Airy pulse initially increases rapidly before a steep decrease and then small oscillation appears. Moreover, the frequency of these oscillations increases with an increase in the N. Also, the temporal position of solitons as depicted in fig. 4 (c) reveals that soliton shifts towards the leading edge as the value of N is increased.

### 3.3 Effect of FOD and Quintic Nonlinearity

Recently, it has been observed that the effect of Kerr nonlinearity can be reduced in metal-dielectric nanocomposite material by adjusting the occupied volume fraction of suspended nanoparticles [29,30]. In that case, nonlinear response related to $\chi^{(5)}$ starts to play the dominant role. Therefore our attention in present section is to study the Airy pulse propagation in a medium where only negative FOD and quintic nonlinearity are present.

The previously defined parameter $\sigma = \gamma_2 P_0^2 T_0^4 / |\beta_4|$ in equation (3) depends upon the input pulse power and denotes the relative strength of FOD and quintic nonlinearity. Therefore, change in the input pulse power is directly related to the change in $\sigma$. Fig. (5) shows the evolution of truncated Airy pulse (a = 0.1) in the temporal and spectral domains respectively for different value of $\sigma$. The progressive supremacy of nonlinear effects can be appreciated for larger value of $\sigma$. When the initial launch peak power of the pulse is relatively small, the nonlinear effects underplay during the propagation and therefore weak soliton shedding is observed. Interestingly, the background radiation of Airy pulse, reported earlier in [16], is not visible in present case because FOD demolishes the accelerating wavefront responsible for this radiation. As delineated in fig. (5), soliton shedding is possible even for σ as small as 0.2, where the energy of the main lobe of the truncated Airy pulse peaks first then it depletes to a soliton structure. The emergent soliton oscillates during the propagation whose periodicity can be managed by changing the launched pulse power. When σ increases further, corresponding peak power and frequency of oscillation of the soliton increases. Moreover, in Fourier domain the spectral broadening is observed.

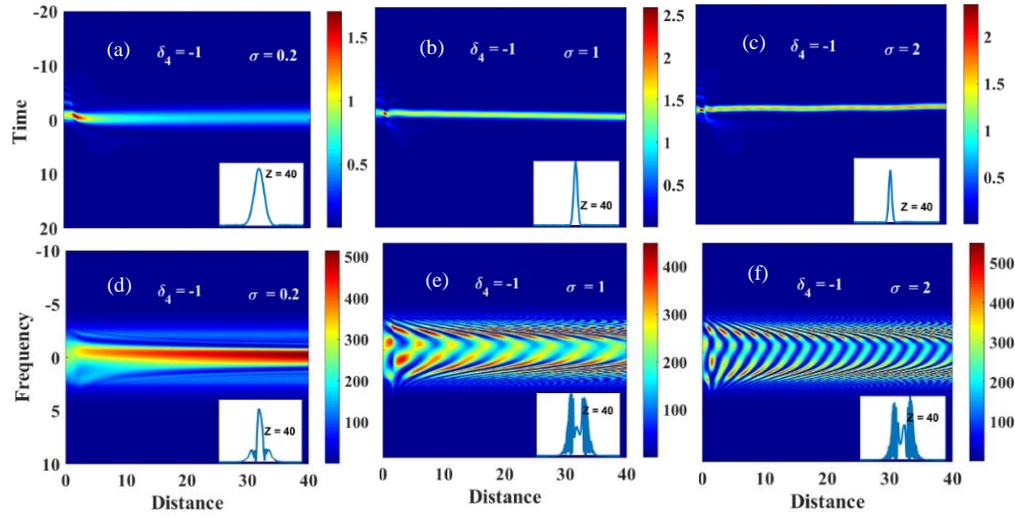

**Fig. 5.** Contour plots of temporal (upper row) and spectral (lower row) Airy pulse evolution under the influence of negative FOD and quintic nonlinearity. **(a), (d)** $\sigma = 0.2$, **(b), (e)** $\sigma = 1$, and **(c), (f)** $\sigma = 2$.

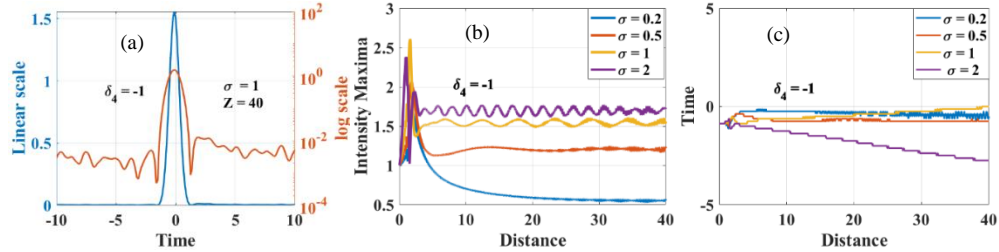

**Fig. 6.** For $\delta_4 = -1$ and a = 0.1 **(a)** output pulse intensity with $\sigma = 1$ **(b)** intensity maximum with propagation distance and **(c)** temporal position of intensity maximum with four different values of quintic nonlinear parameter.

In addition, it is worth mentioning that the intensity of the output pulse on the log scale demonstrates the oscillatory tail at the leading and trailing edges, which can be seen in fig. 6 (a). Also, the maximum intensity of the Airy pulse shown in fig. 6 (b), initially increases before a sharp decrease and thereafter shows small oscillations, and the amplitude of these oscillations increases with an increase in the value of $\sigma$. Temporal position of soliton depicted in fig. 6 (c) demonstrate that emerging soliton can shifts towards either to leading or trailing edge depending on the strength of quintic nonlinearity.

### 3.4 Effect of FOD and Cubic-Quintic Nonlinearity

In this section, we consider a situation when negative FOD and cubic-quintic nonlinearity only play a significant role in the truncated Airy pulse propagation. The temporal and spectral evolution of the pulse is shown in fig. (7). When N = 1 intense soliton shedding is observed in the presence of quintic nonlinearity $\sigma$ and the emergent soliton shifts towards the leading edge with stronger $\sigma$. The inset in figures shows the output pulse shape at Z = 40.

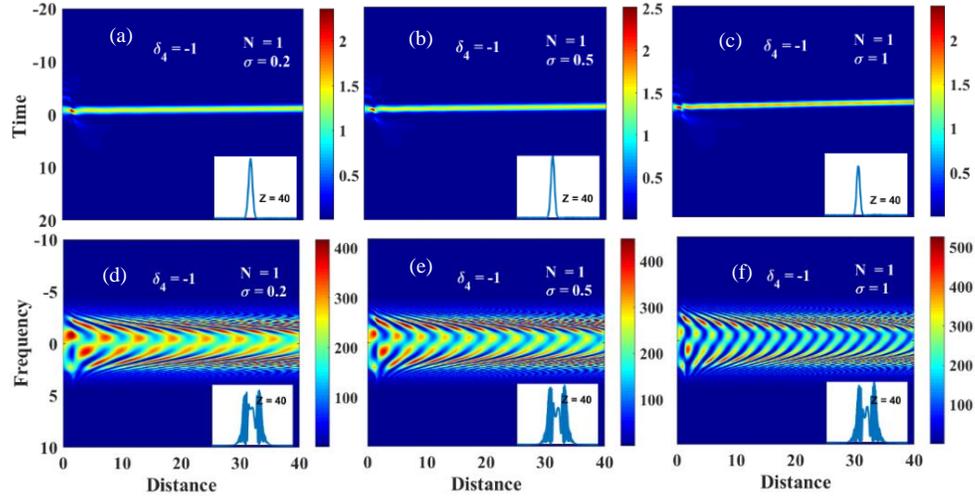

**Fig. 7.** Contour plots of temporal (upper row) and spectral (lower row) Airy pulse evolution under the influence of $\delta_4 = -1$, N = 1. **(a), (d)** $\sigma = 0.2$, **(b), (e)** $\sigma = 0.5$, and **(c), (f)** $\sigma = 1$.

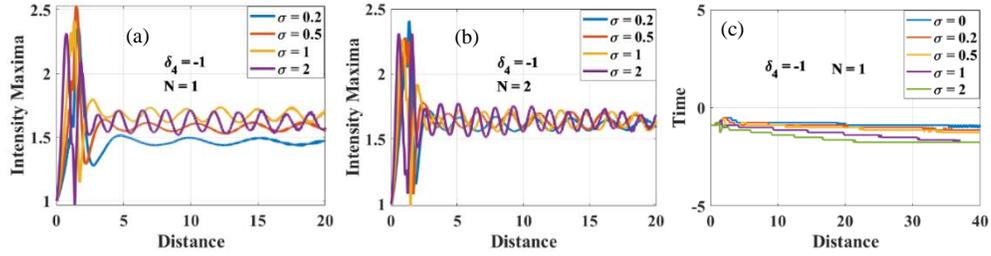

**Fig. 8.** For $\delta_4 = -1$, **(a)** intensity maxima with $N = 1$, **(b)** intensity maxima with $N = 2$, and four different quintic nonlinear parameter **(c)** temporal position for $N = 1$ and different quintic nonlinear parameter.

The intensity maximum of the pulse illustrated in fig. 8(a) oscillates with propagation distance and both the frequency of oscillation and magnitude of the intensity maxima increase with the magnitude of the quintic nonlinearity. For stronger Kerr nonlinearity, the intensity maxima magnitude become almost independent of the quintic nonlinearity as shown in fig. 8(b). The quintic nonlinear parameter influences the temporal shift of the emergent soliton as is shown in fig. 8 (c).

### 3.5 Effect of Anomalous GVD, FOD and Cubic Nonlinearity

Anomalous GVD and SPM are the lowest-order dispersion and nonlinear effects whose combination supports soliton propagation in a medium. Although the soliton-like solution has also been investigated in the regime of anomalous GVD and FOD with cubic nonlinearity [31].

Hence, in this section, we investigate Airy pulse propagation in such a regime and analyse the effect of N. The temporal evolution of the Airy pulse is delineated in fig. 9 (a-c). In fact, soliton shedding is observed in such a case, but the presence of non-zero anomalous GVD results in soliton shedding at relatively larger input power, that is, at larger N value (as compare to fig. (3) case). For N = 1, soliton shedding is observed as can be seen in fig. 9 (a). On increasing the value of N, more energy concentrate in the soliton and soliton shifts towards the leading edge as depicted in fig. 9 (b-c).

Besides, nonlinearity of the medium significantly modifies the spectral evolution of the pulse, as shown in fig. 9 (d-f). The spectral broadening can be noticed for N = 1 in fig. 9 (d). When N is increased further, the multipeak structures of the pulse spectrum becomes complex as delineated in fig. 9 (e-f).

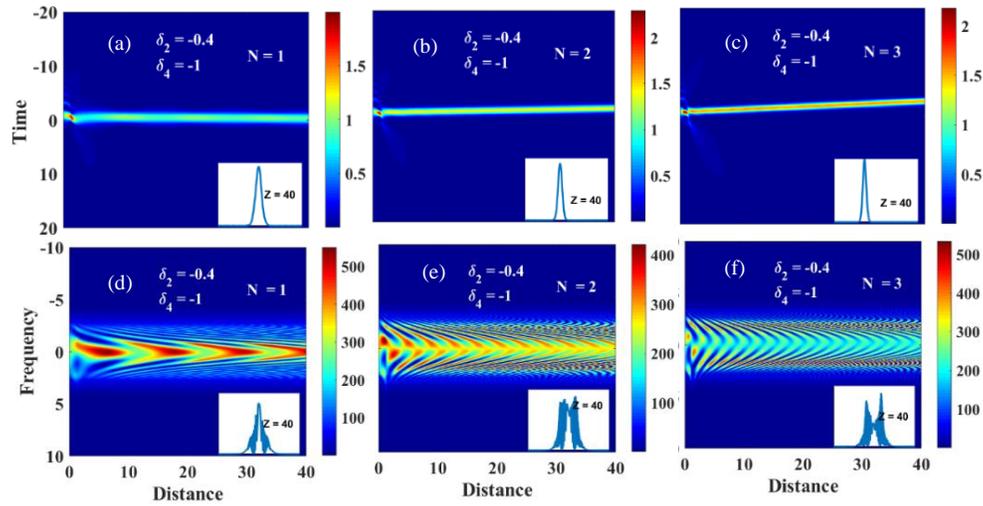

**Fig. 9.** Contour plots of temporal (upper row) and spectral (lower row) Airy pulse evolution under the influence of $\delta_2 = -0.4, \delta_4 = -1$ and cubic nonlinearity. **(a), (d)** $N = 1$, **(b), (e)** $N = 2$, and **(c), (f)** $N = 3$.

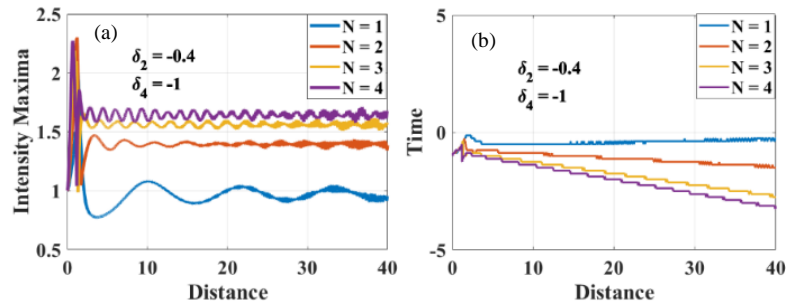

**Fig. 10. (a)** Intensity maxima with $\delta_2 = -0.4 \; \delta_4 = -1$ **(b)**, temporal position with $\delta_2 = -0.4 \; \delta_4 = -1$

Furthermore, we have shown the evolution of intensity maxima of the emergent soliton and corresponding temporal position in fig. 10 (a-b). The intensity maximum oscillates during propagation and for large value of N frequency of oscillations are increased. For the large value of N, temporal shift of the soliton is also increased as clearly shown in fig.10 (b).

### 3.6 Effect of Anomalous GVD, FOD and Cubic-Quintic Nonlinearity

In the presence of anomalous GVD and negative FOD, the combined effect of cubic and quintic nonlinearity leads to soliton shedding which shows very interesting behaviour. The combined effect of nonlinearity facilitates shifting of the emergent soliton in either side of $\tau = 0$, i.e. either on leading or trailing edge depending upon the value of nonlinear parameters. The evolution of the intensity maxima and temporal position of the emergent soliton is shown is fig. 11. For smaller GVD and N = 1, soliton tend to shift towards the trailing edge until the quintic nonlinear parameter reaches 1. Soliton shift towards the leading edge beyond this value of quintic nonlinearity as clearly illustrated in fig. 11 (b). For the N = 2, the soliton shifts towards the leading edge even for the small value of quintic nonlinearity as can be seen in fig. 11 (c).

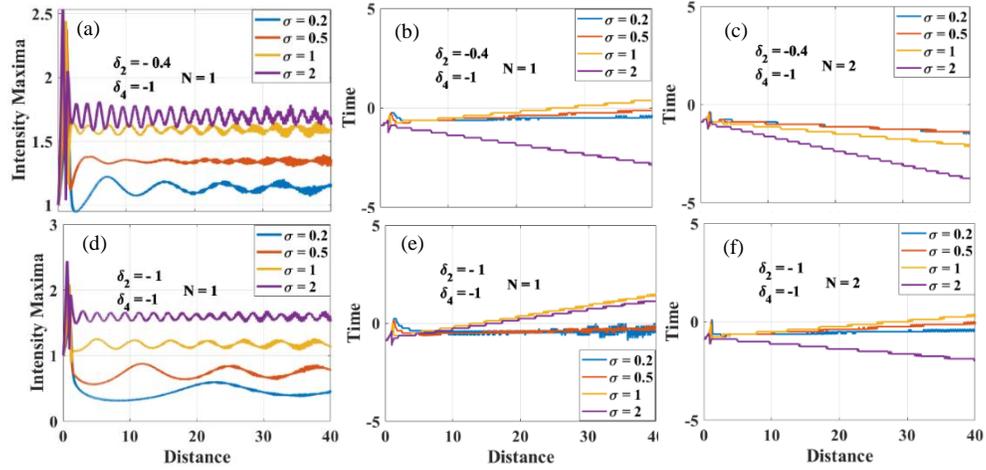

**Fig. 11**. Intensity maxima **(a)** $\delta_2 = -0.4$, $\delta_4 = -1$, $N = 1$, **(b)** temporal position corresponding to (a), **(c)** $\delta_2 = -0.4$, $\delta_4 = -1$, $N = 2$, (upper row), intensity maxima **(d)** $\delta_2 = -1$ $\delta_4 = -1$, $N = 1$, **(e)** temporal position corresponding to (d), **(f)** $\delta_2 = -1$, $\delta_4 = -1$, $N = 2$, (lower row) with different quintic nonlinear parameter.

For larger value of the GVD parameter ($\delta_2 = -1$), we can see that temporal shift increases towards the trailing edge for $N = 1$ and then slows down for $\sigma = 2$ as clearly demonstrated in fig. 11 (e). On increasing the value of N (i.e. N=2), another interesting feature of emergent soliton is observed as can be seen in fig. 11 (f). In the figure, we can see that soliton shifts towards the trailing edge for smaller quintic nonlinearity, while for its larger values soliton shift towards the leading edge.

### 3.7 Effect of Anomalous GVD, FOD and Cubic-Quintic-Septic Nonlinearity

In this section, we investigate the influence of septic nonlinearity on the soliton shedding when cubic quintic nonlinearity is present along with anomalous GVD and FOD. The evolution of the intensity maxima is demonstrated in fig. 12 (a). The figure shows that the soliton intensity oscillates with propagation. A considerable change in the frequency of oscillation and intensity maxima magnitude is noticed once the septic nonlinearity is varied.. The temporal position of the intensity maxima for N=1 is shown in fig. 12 (b). For the N =1, a huge temporal shift is observed on increasing the septic nonlinearity.

For larger values of GVD the evolutions are slightly modified as shown in fig. 12 (c) and fig. 12 (d).

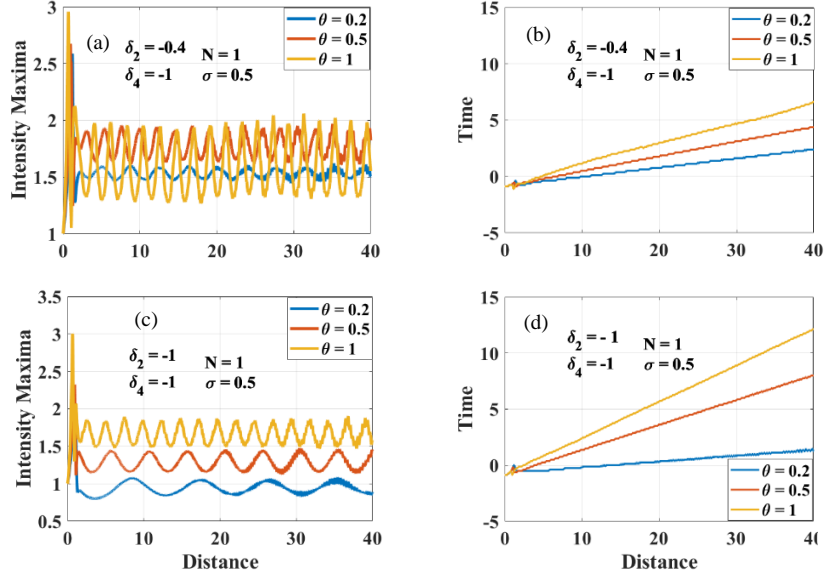

**Fig. 12.** Intensity maxima (first column) and temporal position (second column) under the influence of $\delta_4 = -1$, $N = 1$, $\sigma = 0.5$ **(a), (b)** $\delta_2 = -0.4$ **(c), (d)** $\delta_2 = -0.4$ with different value of septic nonlinearity

## 4. Conclusion

In summary, we have studied the dynamics of truncated Airy pulse propagation in the linear and nonlinear regimes. In the linear regime, oscillatory tail merges into the Airy main lobe and the symmetric shape pulse with increased intensity resulted. In the nonlinear regime, soliton shedding is observed by the interplay between the negative FOD and cubic nonlinearity at the relatively low input power, which exhibits oscillatory tail both at the leading and trailing edges and on increasing the magnitude of the cubic nonlinearity. The resulted soliton shifts towards the leading edge. On the other hand, soliton shedding observed with negative FOD and quintic nonlinearity can shift either towards the leading edge or trailing edge depending upon the strength of the quintic nonlinearity. In addition, propagation is studied where anomalous GVD and negative FOD are present together with cubic quintic and septic nonlinearity and it has been noticed that in the presence of anomalous GVD input pulse power required for soliton shedding is increased. Furthermore, we found that the strength of GVD and quintic and septic nonlinearity play a deciding role in the temporal shift of the emergent soliton. Our results may lead to potential applications in optical communication and signal processing.

**Disclosures.** The authors declare no conflicts of interest.

**Data Availability.** Data underlying the results presented in this paper are not publicly available at this time but may be obtained from the authors upon reasonable request.

**References**

1. M.V. Berry and N.L. Balazs, "Nonspreading wave packets", Am. J. Phys. **47**, 264-267(1979).

2. G. A. Siviloglou, J. Broky, A. Dogariu, and D. N. Christodoulides, " Observation of Accelerating Airy Beam", Phys. Rev. Lett. **99**, 213901 (2007).
3. Agrawal GP. Nonlinear Fiber optics. 4th ed. New York: Academic; 2007.
4. G. A. Siviloglou, J. Broky, A. Dogariu, and D. N. Christodoulides, "Ballistic dynamics of Airy Beam", Opt. Lett. **33**,207 (2008)
5. J. Baumgartl, M. Mazilu, and K. Dholakia, "Optically mediated particle clearing using Airy wavepackets,"Nat. Photonics **2**, 675–678 (2008).
6. P. Zhang, J. Prakash, Z. Zhang, M. S. Mills, N. K. Efremidis, D. N. Christodoulides, and Z. Chen, "Trapping and guiding microparticles with morphing autofocusing Airy beams," Opt. Lett. **36**, 2883–2885 (2011).
7. P. Polynkin, M. Kolesik, and J. Moloney, "Filamentation of femtosecond laser Airy beams in water," Phys. Rev. Lett. **103**, 123902 (2009).
8. P. Panagiotopoulos, D. Abdollahpour, A. Lotti, A. Couairon, D. Faccio, D. G. Papazoglou, and S. Tzortzakis, "Nonlinear propagation dynamics of finite-energy Airy beams," Phys. Rev. A **86**, 013842 (2012).
9. P. Polynkin, M. Kolesik, J. Moloney, G. Siviloglou, and D. N. Christodoulides, "Extreme nonlinear optics with ultra-intense self-bending Airy beams," Opt. Photon. News **21**, 38–43 (2010).
10. C. Ament, P. Polynkin, and J. V. Moloney, "Supercontinuum generation with femtosecond self-healing airy pulses," Phys. Rev. Lett. **107**, 243901 (2011).
11. G. A. Siviloglou, and D. N. Christodoulides, "Accelerating finite energy Airy beams", Opt. Lett. **32**, 979 (2007)
12. T. Latychevskaia, D. Schachtler, and H. W. Fink, " Creating Airy Beam employing a transmissive spatial light modulator", App. Opt. **55** (22) (2016)
13. Thawatchai Mayteevarunyoo, and Boris A. Malomed, "Generation of $\chi^2$ solitons from the Airy wave through the parametric instability", Opt. Lett. **40,** 4947-4950 (2015)
14. Thawatchai Mayteevarunyoo, and Boris A. Malomed, "Two-dimensional $\chi^2$ solitons generated by the down conversion of Airy waves", Opt. Lett. **41**, 2919-2922 (2016)
15. Thawatchai Mayteevarunyoo, and Boris A. Malomed, "The interaction of Airy wave and solitons in a three wave system", J. Opt. **19** (2017)
16. Y. Fattal, A. Rudnick, and D. M. Marom, "Soliton shedding from airy pulses in Kerr media," Opt. Express **19**, 17298–17307 (2011).
17. Y. Hu, M. Li, D. Bongiovanni, M. Clerici, J. Yao, Z. Chen, J. Azaña, and R. Morandotti, "Spectrum to distance mapping via nonlinear Airy pulses," Opt. Lett. **38**, 380–382 (2013).
18. R. Driben, Y. Hu, Z. Chen, B. A. Malomed, and R. Morandotti, "Inversion and tight focusing of Airy pulses under the action of third-order dispersion", Opt. Lett. **38**, 2499–2501 (2013).
19. L. F. Zhang, J. G. Zhang, Y. Chen, A. L. Liu, and G. C. Liu, "Dynamic propagation of finite-energy Airy pulses in the presence of higher-order effects", J. Opt. Soc. Am. B **31**, 889–897 (2014).
20. Martin Djoko, T. C. Kofane, "The cubic–quintic–septic complex Ginzburg–Landau equation formulation of optical pulse propagation in 3D doped Kerr media with higher-order dispersions", Opt. Comm. 416, 190-201 (2018)
21. Anders Hook and Magnus Karlsson, "Ultrashort solitons at the minimum-dispersion wavelength: effects of fourth-order dispersion", Opt. Lett. **18**, 1388, (1993)
22. Ivan P. Christov, Margaret M. Murnane, Henry C. Kapteyn, Jianping Zhou, and Chung-Po Huang, "Fourth-order dispersion-limited solitary pulses", Opt. Lett. **19**, 1465 (1994)
23. Michel Piche, Jean-Francois Cormier, and Xiaonong Zhu, "Bright optical soliton in the presence of fourth-order dispersion", Opt. Lett, **21**, 845 (1996)
24. Sarma AK., "Solitary wave solutions of higher-order NLSE with Raman and self-steepening effect in a cubic–quintic–septic medium.", Commun Nonlinear Sci Numer Simul **14**, 3215–3219 (2009).
25. Kevin K. K. Tam, Tristram J. Alexander, Andrea Blanco-Redondo, and C. Martijn De Sterke, "Stationary and dynamical properties of pure-quartic solitons", Opt. Lett. **44**, 3306 (2019)
26. Kevin K. K. Tam, Tristram J. Alexander, Andrea Blanco-Redondo and C. Martijn de Sterke, "Generalized dispersion Kerr solitons", Phy. Rev. A **101**, 043822 (2020)
27. Andrea Blanco-Redondo , C. Martijn de Sterke , J.E. Sipe, Thomas F. Krauss, Benjamin J. Eggleton and Chad Husko, "Pure-quartic solitons", Nat. Commun. **7**:10427 doi: 10.1038/ncomms10427 (2016)
28. A. Mohamodou, C. G. LatchioTiofack, and Timoléon C. Kofané, "Wave train generation of solitons in system with higher-order nonlinearities", Phy. Rev. E **82**, 016601 (2010).
29. Albert S. Reyna and Cid B. de Araujo, "Nonlinearity management of photonic composites and observation of spatial-modulation instability due to quintic nonlinearity", Phy. Rev. A **89**, 063803 (2014)
30. Albert S. Reyna and Cid B. de Araújo, "Spatial phase modulation due to quintic and septic nonlinearities in metal colloids", Opt. Exp. **22**, 22456 (2014)
31. N.N. Akhmediev, A.V. Buryak, M. Karlsson, "Radiationless optical solitons with oscillating tails", Opt. Comm. **110**, 540-544 (1994)